\documentclass[12pt]{article}
\usepackage{fullpage}
\usepackage{epsfig}
\usepackage{epsf}
\usepackage{amsmath}
\usepackage[footnotesize]{caption}
\usepackage{setspace} 
\usepackage{natbib}
\usepackage{rotating}
\usepackage[displaymath]{lineno}
\usepackage{geometry}
\usepackage{setspace}
\usepackage{times}
\usepackage{enumerate}

\newcommand{\dd}[2]{\frac{d#1}{d#2}}

\newcommand{\mr}[1]{\mathrm{#1}}
\newcommand{\ith}[1]{$i^{\mr{th}}$}
\newcommand{\degC}{$^\circ$C}

\begin{document}

\begin{centering}
{\bf Understanding uncertainty in temperature effects on vector-borne disease: A Bayesian approach}\\
\bigskip
Leah R. Johnson$^{*1,2}$, Tal Ben-Horin$^{3,4}$, Kevin D. Lafferty$^{5,6}$, Amy McNally$^{7}$, Erin Mordecai$^{3,8}$, Krijn P. Paaijmans$^{9}$, Samraat Pawar$^{1,10}$, Sadie J. Ryan$^{11,12,13}$
  \\
\bigskip

\footnotesize 1-Ecology and Evolution, University of Chicago; 
\footnotesize 2-Integrative Biology, University of South Florida;\\
\footnotesize 3-Ecology, Evolution, and Marine Biology, UC Santa Barbara; 4-Marine and Coastal Sciences, Rutgers University;
\footnotesize 5-Western Ecological Research Center, US Geological Survey;  6-Marine Science Institute, UC Santa Barbara;\\
 \footnotesize 7-Geography Department, UC Santa Barbara;  8-Biology, UNC Chapel Hill; \\
 \footnotesize 9-Barcelona Centre for International Health Research, Universitat de Barcelona; \\
\footnotesize 10-Department of Life Sciences, Imperial College;  11- Environmental and Forest Biology, SUNY ESF\\
\footnotesize 12-Center for Global Health and Translational Science, SUNY UMU; \\ 
\footnotesize 13 - School of Life Sciences College of Agriculture, Engineering, and Science, \\
\footnotesize University of KwaZulu-Natal, Durban, South Africa\\
\end{centering}

\linenumbers 
\modulolinenumbers[2]
 \doublespace

\raggedright
\noindent {\bf ABSTRACT:} Extrinsic environmental factors influence the distribution and population dynamics of many organisms, including insects that are of concern for human health and agriculture. This is particularly true for vector-borne infectious diseases, like malaria, which is a major source of morbidity and mortality in humans. Understanding the mechanistic links between environment and population processes for these diseases is key to predicting the consequences of climate change on transmission and for developing effective interventions. An important measure of the intensity of disease transmission is the reproductive number $R_0$. However, understanding the mechanisms linking $R_0$ and temperature, an environmental factor driving disease risk, 
can be challenging because the data available for parameterization are often poor. To address this we show how a Bayesian approach can help identify critical uncertainties in components of $R_0$ and how this uncertainty is propagated into the estimate of $R_0$.  Most notably, we find that different parameters dominate the uncertainty at different temperature regimes: bite rate from 15-25$^\circ$ C;  fecundity across all temperatures, but especially $\sim$25-32$^\circ$ C; mortality from 20-30$^\circ$ C; parasite development rate at $\sim$15-16$^\circ$C and again at $\sim$33-35$^\circ$C. 
Focusing empirical studies on these parameters and corresponding temperature ranges would be the most efficient way to improve estimates of $R_0$. 
While we focus on malaria, our methods apply to improving process-based models more generally, including epidemiological, physiological niche, and species distribution models.


\clearpage
\section{Introduction}

Malaria is a vector-borne disease that is a major source of illness and mortality in humans, especially in developing countries. Like many vector-borne diseases, the dynamics of malaria are greatly influenced by extrinsic environmental factors such as temperature and rainfall.  As climate changes over time, the distribution of both epidemic and endemic malaria will likely change as well, presenting new challenges for control. A better understanding of how the dynamics of malaria depend on environmental factors will be vital for understanding and planning for shifts in malaria incidence. 

Various approaches have been used to try to understand the question of how environmental change is likely to impact the prevalence and distribution of malaria \citep[reviewed in][]{guerra_mapping_2007,johnson_mapping_2014}. Many of these models can be classified as niche or species distribution models, and they seek to link climate factors to observations of the prevalence of vectors, parasites, or disease occurrence. For mechanistic versions of these models (in contrast to geographical correlation models), it is necessary to understand how the vital rates of all players in disease transmission respond to the environment. Temperature strongly influences vital rates, particularly in ectotherms, and its effects can be measured under laboratory conditions. Despite this basic premise and our reasonable knowledge about thermal physiology, data on responses of vital rates to temperature are not widely available. 
Even for species that have been well studied, like malarial parasites and their mosquito vectors, the quality and quantity of the data are uneven across traits and temperatures. The paucity of data compromises the quality of model predictions, such as the range of temperatures that are conducive to disease transmission. Moreover, the sensitivity of model predictions to errors in empirical estimates is not well known. Here, we develop methods for estimating sensitivity of model outputs to model inputs, focusing on the effect of temperature on malaria transmission.

An important and simple measure of transmission, the reproductive number, $R_0$, is often used in disease studies as it is related to both how quickly a disease can spread in a na\"{i}ve population and to the level of prevalence (the proportion of individuals that have been infected) for endemic diseases \citep{keeling_modeling_2008}. Recently, approaches have been developed to model how $R_0$ depends on temperature by incorporating thermal responses of traits underlying $R_0$, such as mosquito and parasite development rates \citep[]{mordecai_optimal_2013,molnar_metabolic_2013}. For malaria, the
method involves the use of laboratory data collected on the temperature dependence of all
components of $R_0$ that depend upon parasite or vector physiology --- 
temperature response curves fitted to each component are incorporated into
the $R_0$ equation to find the overall thermal dependence of transmission. Including these physiologically based thermal responses produces predictions of transmission that are more inline with observed incidence patterns than previous models that do not incorporate these detailed physiological responses, even without accounting for rainfall \citep[]{mordecai_optimal_2013}. Although many other factors, such as key control measures, may better predict malaria morbidity and mortality, this kind of relatively simple modeling is a promising first step in prioritizing global health policy to respond to broad changes in the spread and intensification of infectious diseases \citep{altizer_climate_2013}.

Empirical research on the factors or components that determine $R_0$ is costly. Thus, it is important to direct future research towards aspects that will result in the greatest reduction in uncertainty in $R_0$ overall, and thus improve our predictions of changes in future transmission the most. Currently, the laboratory data necessary to understand the temperature dependence of the components that determine the response of  $R_0$ to temperature are often limited. Available data leave substantial uncertainty about the relationship between each component of $R_0$ and temperature, especially at the temperatures that are marginal for transmission. Thus, we anticipate considerable uncertainty in how $R_0$ varies with temperature. By better understanding all of the sources of uncertainty we can prioritize laboratory studies more efficiently and design effective intervention strategies \citep{elderd:2006,merl:2009}. \citet{mordecai_optimal_2013} addressed this issue to some degree by performing a sensitivity analysis
by perturbing the $R_0$ components with respect to temperature. However, this
kind of simple, single-parameter local sensitivity analysis does not allow a full understanding of either the uncertainty in components or in $R_0$ overall. Further, additional data on components of $R_0$ for closely related species or less well-controlled experiments are often available. These additional data, even if not ideal for fitting the final models directly, can be informative.

Here we use a Bayesian approach \citep{clark}  to understand the full range of uncertainty in the thermal
response of malarial $R_0$. 
The focus of a Bayesian analysis is the posterior distribution: i.e., the probability that the parameters have some value given the data. This is obtained by combining a likelihood (the probability of observing the data given parameters with particular values) and a prior distribution (the assumed probability that the parameters have some values independent of the observed data) using Bayes rule. A full discussion of the Bayesian approach can be found elsewhere \citep[e.g.][]{clark}. 
A Bayesian approach allows us to incorporate prior knowledge about the various components of $R_0$, for instance, by using data from related species in the inference procedure. This is especially useful in applications that rely heavily on sparse data, such as the one explored here. 

We are interested in two primary aspects of the relationship between transmission and temperature: (1) which temperatures prevent transmission? and (2) which temperatures promote transmission? Earlier work on temperature and disease transmission in general, and for malaria in particular, has produced mixed results, in part because the impact of temperature on preventing transmission (as opposed to promoting it) is often ignored \citep{rohr_frontiers_2011,hay_climate_2002,siraj_altitudinal_2014,gething_climate_2010}. 
We use a Bayesian approach to explore the uncertainty and sensitivity of these two transmission outcomes -- prevention and promotion -- to mosquito and parasite traits.

We begin by introducing the 
$R_0$ model and its components, the potential thermal responses for all its components, and
the available data on these thermal responses. We then introduce the data model,
initial ``uninformative'' priors, and our overall methodology. We then step through a series of uncertainty and sensitivity analyses, together with the results for each analysis. This is followed by a discussion of how the approach taken compares to more classical analyses, and the implications of the results. 

\vspace{-0.5cm}
\section{Data, Models, and Methods}
The standard model of malaria transmission by a vector is the Ross-McDonald
model \citep{macdonald_analysis_1952}, from which the reproductive number
$R_0$ is derived. $R_0$ determines the dynamical threshold for disease
transmission, and is defined as the average number of secondary infections
caused by a single infected individual in an entirely susceptible population.
It specifies the relationships of parameters in the model that are
required for an infection to spread within a population ($R_0>1$) as opposed to
dying out ($R_0<1$). The most widely used formulation for malarial
$R_0$ \citep{dietz_estimation_1993} is
\begin{equation}
R_0 = \sqrt{\frac{M}{Nr} \frac{a^2bc\exp{(-\mu/PDR)}}{\mu}}, \label{eq:R0}
\end{equation}
where $M$ is the density of mosquitoes, $a$ is the bite rate, $bc$ is vector competence, $\mu$ is the mortality rate of adult mosquitoes, $PDR$ is the parasite development rate (1/$EIP$, the extrinsic incubation period of the parasite), $N$ is the human density, and $r$ is the human recovery rate. Most of these model components are directly measurable or are closely related to quantities or traits that can be observed \citep{mordecai_optimal_2013}. Following \citet{mordecai_optimal_2013}, we assume that the expected mosquito density is given by:
\begin{equation}
M=\frac{EFD p_{EA} MDR}{\mu^2}, \label{eq:M}
\end{equation}
where $EFD$ is number of eggs produced per female per day, $p_{EA}$ is the probability that an egg will hatch and the larvae will survive to the adult stage, and $MDR$ is the mosquito development rate. The parameters that jointly define $R_0$ and $M$ are summarized in Table \ref{tb:params}, and throughout this paper we refer to these as ``components of $R_0$''.

Virtually all physiological traits in ectotherms exhibit unimodal temperature
responses, i.e., they have an optimal temperature at which the trait is maximized,
and declines on either side \citep[e.g.,][]{amarasekare_framework_2012,dell_systematic_2011,angilletta_thermal_2009}. 
However, the exact functional form of the unimodal response is still
under debate, especially because it is known to vary with the type of trait
\citep{dell_systematic_2011,mordecai_optimal_2013}.
Therefore, as in \citet{mordecai_optimal_2013}, we determined
the appropriate thermal-response model for each component trait by fitting
candidate functional forms: quadratic for symmetric responses and Briere for asymmetric (see Section \ref{s:likelihood} and Figure \ref{f:allcomps}). These were chosen as they are among the simplest functional forms that exhibit the desired unimodal behavior.  

All analyses were conducted in {\sf R} \citep{cranR} with Markov chain Monte Carlo (MCMC) implemented in {\tt rjags}/JAGS (Just Another Gibbs Sampler, \citep{plummer2003jags,rjags}). Computer code for all analyses are included in the Supplementary Files. 

\vspace{-0.5cm} 
\subsection{{\small Data}}

We use two sets of data in our analysis --- the ``main'' dataset contains the focal data for the thermal responses for the components that make up $R_0$ (Table \ref{tb:params}), and a ``prior'' dataset used to elicit priors for our Bayesian
analysis (both sets included in the Supp.~Files; for sources see Supp.~App.~A, Table 1). Ideally, our main data set would exclusively comprise laboratory data on {\it Plasmodium falciparum}, the causative agent of the majority of tropical malaria, and its primary vector {\it Anopheles gambiae}, held at constant temperature for all components. These data were available for only three traits: mosquito development rate (MDR), egg to adult survival ($ p_{EA}$), and adult mosquito mortality rate ($\mu$). For other traits ideal data are unavailable \citep{mordecai_optimal_2013}, and instead we used data from related species collected under appropriate laboratory conditions. More specifically, we prioritized data collected in the laboratory at constant temperatures. For mosquito and parasite traits, we prioritized based on the relative to efficacy of transmission and severity of disease in humans. Thus, for mosquito traits, we prioritized data for {\it An.~gambiae}, followed by other anophelene species, and finally for {\it Aedes} species. For parasite traits, {\it P.~falciparum} was prioritized, followed by {\it P.~vivax}. When possible, only a single mosquito or parasite species was used for an individual trait in the main data. For our prior data set, our conditions for inclusion were more flexible. Although data on related species held in the lab at constant temperature were preferred, we also allowed more distantly related species, or less controlled (variable temperature) experiments. These data are distinct from the main data set and were used, along with expert opinion, to elicit informative priors for the parameters of the unimodal temperature responses for each $R_0$ component. 

\vspace{-0.5cm}
\subsection{{\small Approach, Likelihoods, and Priors}}
We fitted the thermal response of each component of $R_0$ (Table \ref{tb:params}) to independent data using a Bayesian approach to obtain the posterior distribution for the parameters that describe each response (and thus the posterior distribution of the response itself), as well as for $R_0$ overall. Inference in the Bayesian framework proceeds in three steps. First, a likelihood is defined for
each type of data. Second, appropriate prior distributions are determined. Third,
samples from the posterior distribution of the parameters, given the
data, are obtained via Markov-chain Monte Carlo (MCMC). We used this procedure first for the prior data and for the main data, assuming uninformative priors in both cases (Section \ref{s:priors} and
Supp.~App.~A Section 2). Next, the posterior distributions obtained from analyzing the prior data were used to build informative priors and the inference procedure was repeated for the main data using these informative priors. We then compared the resulting posterior distributions obtained using the uninformative and informative priors. Further, we calculated $R_0$ with both sets of results, and compared these. This gave an indication of the sensitivity of the individual components and of $R_0$ to the choice of prior. We followed this with further
sensitivity and uncertainty analyses (Section \ref{s:sensitivity}).

\vspace{-0.35cm}
\subsubsection{{\small Likelihoods}}\label{s:likelihood}

We assumed functional forms for each component based on our previous work \citep{mordecai_optimal_2013} and on the types of functional forms (unimodal and frequently asymmetric) that are typical for similar traits in other arthropod species \citep{dell_systematic_2011}. More specifically, we used either a quadratic (symmetric) or Briere (asymmetric) function, depending on the component. At any given temperature, the mean response should be determined by this functional form. Further, all model components are, by definition, greater than or equal to 0. Thus, we chose to use a truncated normal distribution, with mean parameter (usually denoted by $\mu$) given by the appropriate functional form (i.e., Briere or quadratic), as the likelihood for the data for most of the components of $R_0$. For all of the components we examined, the lower truncation limit was at zero. Most components can take any value greater or equal to zero. Thus, for most of our data $Y_i | 0<Y_i<b \stackrel{iid}{\sim} \mr{N}(\mu=f(T_i), \sigma^2)$, where $b$ is the upper truncation limit (either 0 or $\infty$), $f(T_i)$ is the Briere or quadratic temperature response, and $iid$ indicates that the data are independently and identically distributed. However, two components, the vector competence ($bc$) and the egg to adult survival ($p_{EA}$), are probabilities and are thus constrained to be between zero and one. For these components, we would ideally have the actual numbers of successes and total numbers of observations so that the more appropriate binomial model could be used. These data were indeed available for vector competence, and so were modeled with a binomial likelihood, i.e. $Y_i \stackrel{iid}{\sim} \mr{Bin}(n, p=f(T_i))$, where $n$ is the number of total observations, of which $Y$ were successes, and the probability, $p$ of a success at a particular temperature, $f(T_i)$, is either Briere or quadratic. For egg to adult survival the raw data were not available so we used a normal distribution truncated at zero and one to model the proportion of eggs that successfully mature to the adult stage. This choice keeps calculations simple, allows straightforward implementation of biologically based priors, and has shape properties that are more appropriate for these data than alternatives such as a beta distribution.


\vspace{-0.35cm}
\subsubsection{{\small Priors}}\label{s:priors}

We began by defining a set of default priors for all parameters that are chosen to be relatively ``uninformative''. That is, these priors were designed to constrain parameter values to be biologically reasonable, but to otherwise  provide wide, reasonably even support across potential parameter values. In particular, for our default priors, we assumed that the maximum temperature at which a unimodal, hump-shaped component goes to zero is $45^\circ$C, and the minimum temperature should be $0^\circ$C, as these temperatures are generally lethal to mosquitoes. This upper limit is slightly higher than some observed upper lethal limits, which are closer to $40^\circ$C \citep{bayoh_studies_2001,bayoh_effect_2003,lardeux_physiological_2008}. We chose the higher, conservative, limit to allow a broader range of temperatures for which the data could inform the posterior distributions. Each of the concave-down (or hump-shaped) curves have a parameter that describes the temperatures at which trait goes to zero, notated as $T_0$ and $T_m$ for the lower and upper limits, respectively. Since we required $T_0 < T_m$ we specified non-overlapping priors for these parameters. For the concave-up quadratic we chose priors that limited the quadratic curves to those that are concave up and in the appropriate quadrant.  We set the priors on other parameters (including the precision parameter, $\tau=1/\sigma$, in the normal distribution) to be diffuse, i.e., to have wide support. Details can be found in Supp.~App.~A, Sections 2 and 3. In all cases we examined the sensitivity of the posterior distributions to the priors. 

\vspace{-0.35cm}
\section{Uncertainty and Sensitivity Analyses and Results}\label{s:sensitivity}

Our uncertainty and sensitivity analyses consisted of multiple parts. First, we addressed sources of uncertainty in our analysis, to understand the expected response of $R_0$ and its components to temperature, and the range of responses that are supported by data. This is similar to global sensitivity analysis for the components and $R_0$. Our measure of uncertainty for each analysis is the 95\% highest probability density (HPD) interval which gives the range of a parameter or response corresponding to a central area containing 95\% of the probability. Second, we compared how the uncertainty in $R_0$ overall depends on the uncertainty in its components, using a variant of local sensitivity analysis, and comparing the results to those obtained for the global-style analysis. Third, we addressed how sensitive $R_0$ is to temperature and to its components, as well as the uncertainty in these relationships. Here, sensitivity is the amount by which $R_0$ changes when temperature changes, and is given by  the derivative of $R_0$ with respect to $T$. As with $R_0$ itself, the uncertainty in this sensitivity is expressed in terms of the 95\% HPD intervals. How the Bayesian and classical approaches to these analyses compare is addressed further in the discussion.

\vspace{-0.5cm}
\subsection{{\small Uncertainty in the components of $R_0$}}
\noindent {\bf How uncertain are the responses of the mosquito and parasite traits to temperature, and how does this depend on the prior information included in the analysis?} \\

To answer this question we made three sets of comparisons.  First, we qualitatively examined the impact of adjusting the default priors for each parameter on the inference for individual components (with both the main and prior data). Second, we compared the posterior distributions for individual components of the main data set  obtained with default and informative priors. In this case, the informative priors are generated from the posterior distributions obtained from fitting the prior data to elicit informative priors. Third, we examined the impact of using an alternative functional response for the vector competence term, for which there is both relatively little data and little {\em a priori} support for a particular functional form. 

The impact of including the informative priors varied between components, in some cases decreasing uncertainty (smaller HPD intervals around the mean -- $bc$, $p_{EA}$,  $EFD$, $PDR$), sometimes increasing uncertainty (larger HPD intervals around the mean -- $a$, $MDR$), or having little impact on the posterior (no change in HPD intervals -- $\mu$) (Supp.~App.~B, Section 1). Across components, modifying priors on the lower and upper temperature limits of the responses ($T_0$ and $T_m$, respectively) had the greatest impact on the posterior distributions of the temperature responses overall. The upper limit of 45$^\circ$C on $T_m$ for the default and some of the informative priors was important for components for which no high temperature data were available, such as the bite rate, $a$. Full results for each component, including inferences with both types of priors, and a comparison of the marginal posterior distributions of parameters with their priors, are included in Supp.~App.~B. 

\begin{figure}[h!] \footnotesize
\centering
\begin{tabular}{ccc}
MDR & PDR & $a$ \\
\includegraphics[trim=20 0 0 50, scale=0.315]{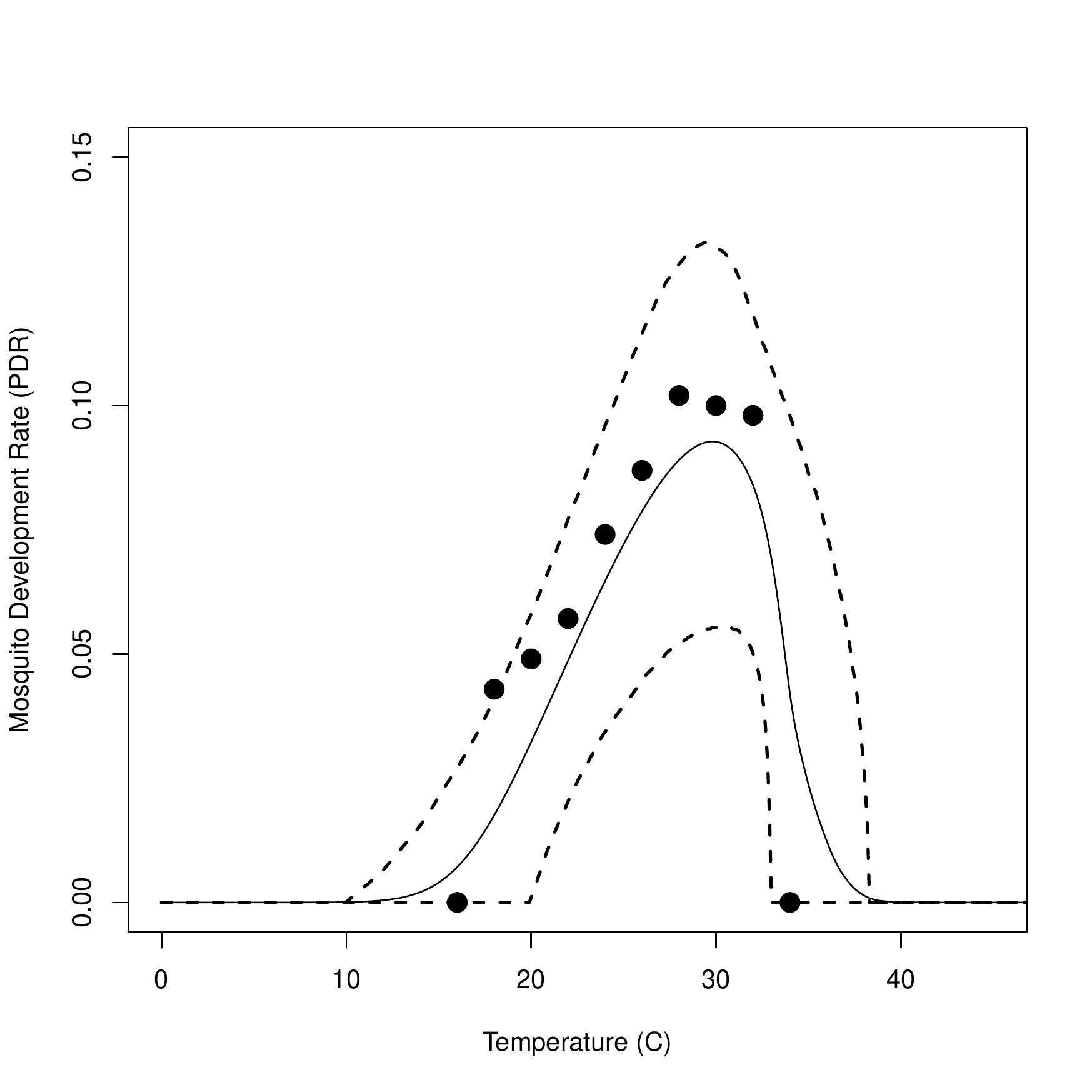}&
\includegraphics[trim=20 0 0 50, scale=0.315]{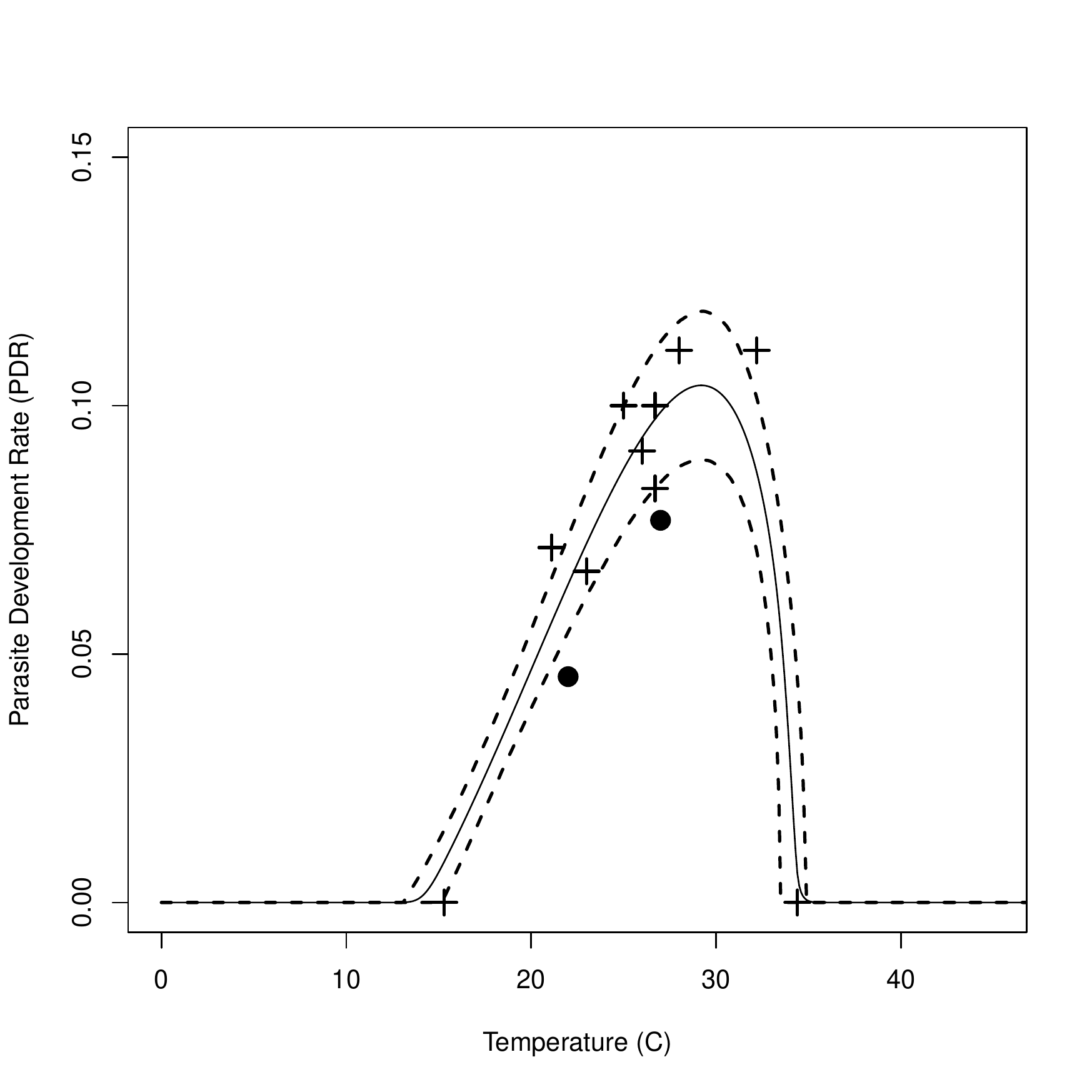} &
\includegraphics[trim=20 0 0 50, scale=0.315]{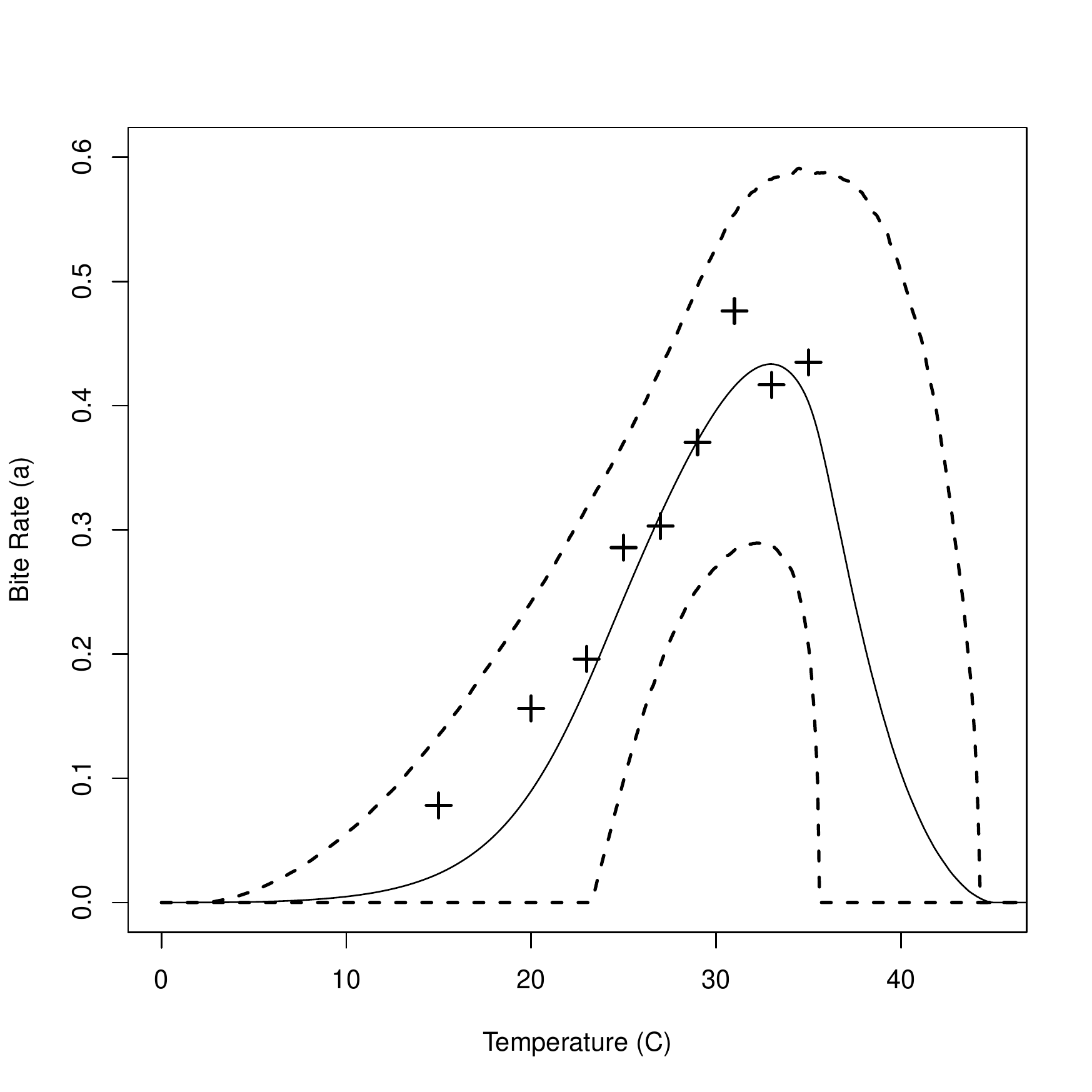}\\
&& \\
$bc$ & $p_{EA}$ & EFD \\
\includegraphics[trim=20 0 0 50, scale=0.315]{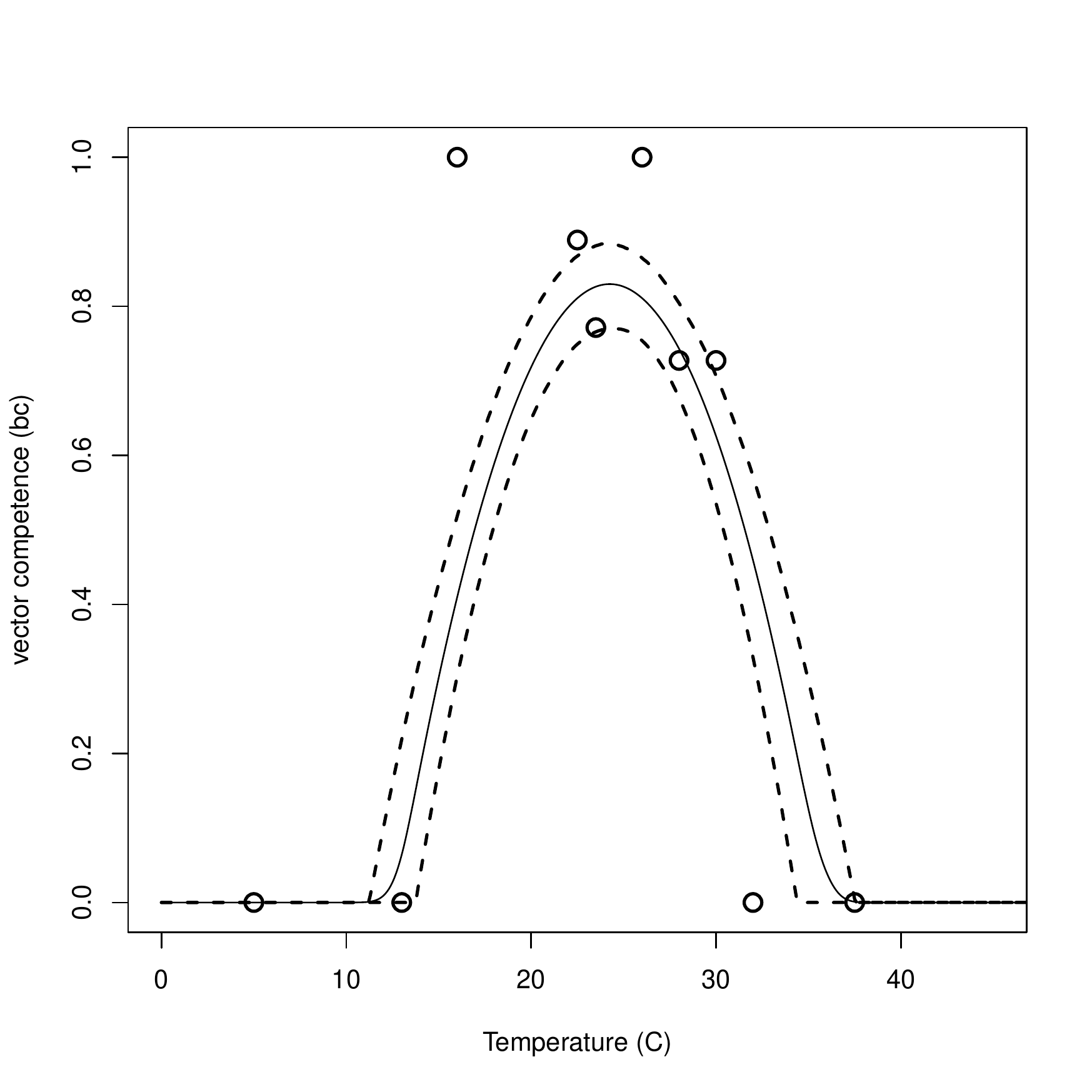} &
\includegraphics[trim=20 0 0 50, scale=0.315]{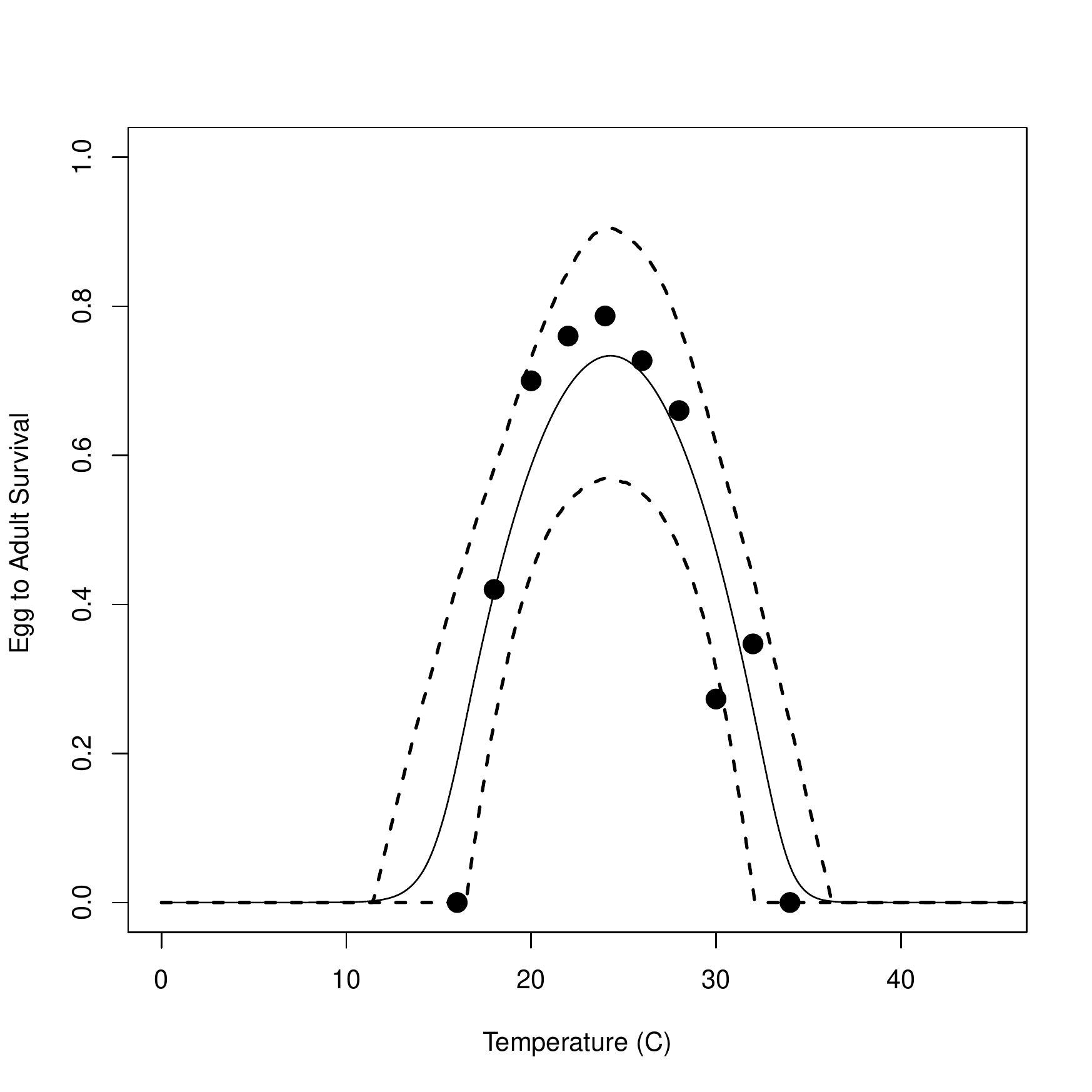} &
\includegraphics[trim=20 0 0 50, scale=0.315]{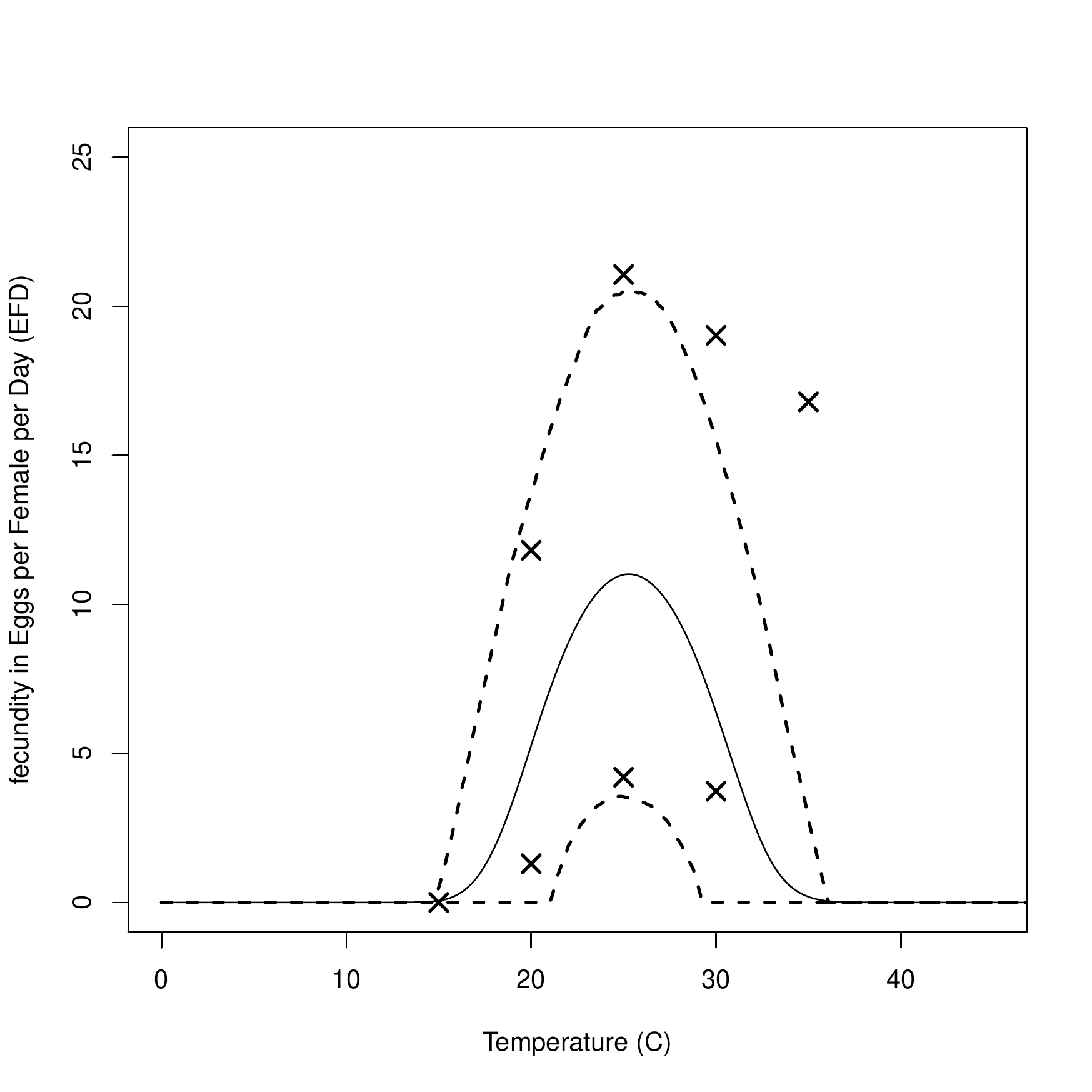}\\
&&\\
& $\mu$ & \\
 & \includegraphics[trim=20 0 0 50, scale=0.315]{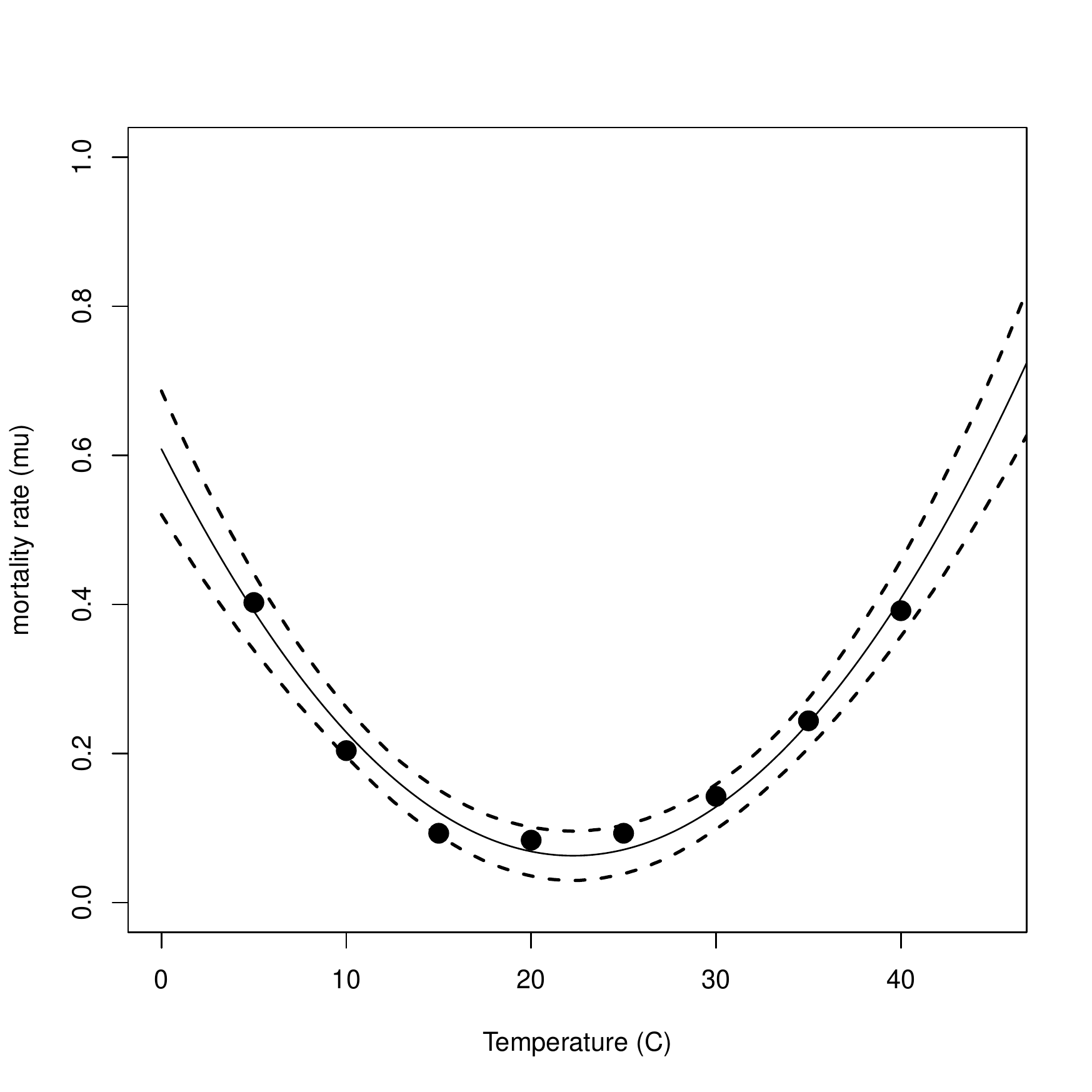}& 
\\
\end{tabular}
\caption{Posterior mean (solid line) and 95\% credible interval (dashed lines) of the thermal responses for all components of $R_0$, with informative priors together with the main data. Traits modeled with a Briere thermal response  ($cT(T-T_0)\sqrt{(T_m-T)}$) are grouped in the top row, concave-down quadratic ($f(T)=a(T-T_0)(T-T_m)$) in the middle row, and concave-up quadratic ($aT^2+bT+c$) in the bottom row. Data symbols correspond to the species of mosquito or parasite used for the analysis. $\bullet$: {\it An.~gambiae} or {\it P.~falciparum} in {\it An.~gambiae}; $+$: other Anophelene species or {\it P.~falciparum} in other Anophelene species; $\times$: {\it Aedes} species; $\circ$: {\it P.~vivax} in other Anophelene species.}
\label{f:allcomps}
\end{figure}

In Figure \ref{f:allcomps} we show the posterior mean and 95\% HPD interval around this mean (summarizing the extent of our uncertainty around this response) for all the components when informative priors were used. Some interesting patterns emerged when we compared across components. First, for all components modeled with a Briere function (top row of Figure \ref{f:allcomps}), the low temperature limit (the temperature below which the trait is zero) was less certain than the upper temperature limit (although this difference was small for PDR). This was partly due to the nature of the functional form -- i.e., it goes to zero more quickly at high temperatures than it does at low temperatures. However, it also reflected that there were often fewer data available across lower temperatures than high temperatures in the main and prior data together. This pattern of uncertainty at the limits was not found for the concave down quadratic responses (middle row, Figure \ref{f:allcomps}). Instead, in some cases, the upper limit was less certain than the lower. This indicates that the temperature resolution for experiments needed to pin down the responses may depend on the type of response (asymmetric vs.~symmetric) that a trait exhibits.

For most of the components explored, either the data gave a strong indication of whether a symmetric or asymmetric response was appropriate \citep{mordecai_optimal_2013}, or there were biophysical reasons why we expected a response to be asymmetric or symmetric {\it a priori} \citep{angilletta_thermal_2009,dell_systematic_2011}. 
 However, vector competence (a compound trait) was ambiguous. Thus, we fit both a quadratic and Briere function for this component. Both fit quite well, and thus the impact of fits using both functional forms on the uncertainty in $R_0$ was addressed in the subsequent analysis.

\vspace{-0.5cm}
\subsection{{\small Overall uncertainty in $R_0$}}
\noindent {\bf How uncertain is the response of the basic respoductive rate, $R_0$, to temperature (due to uncertainty in all components), and how does this depend on the prior information included in the analysis?} \\

To answer this question we made three comparisons.  First, we compared the posterior distributions of $R_0$ under default and informative priors for the components, looking at the overall uncertainty (95\% HPD interval) of the full response curve when all components were allowed to vary according to their posterior distributions. Second, we examined the HPD intervals of three important summaries of $R_0$: minimum (low temperature transmission limit), maximum (high temperature transmission limit), and peak (temperature at maximal transmission) $R_0$. Third, we examined the impact of the two functional responses for the vector competence term on the posterior distribution of $R_0$. This analysis shows the overall uncertainty around (1) which temperatures prevent transmission (low and high temperature transmission limits) and (2) which temperatures promote transmission (peak temperature).

\begin{figure}[h!]
\centering
\begin{tabular}{c}
\includegraphics[trim=0 0 0 0, scale=0.5]{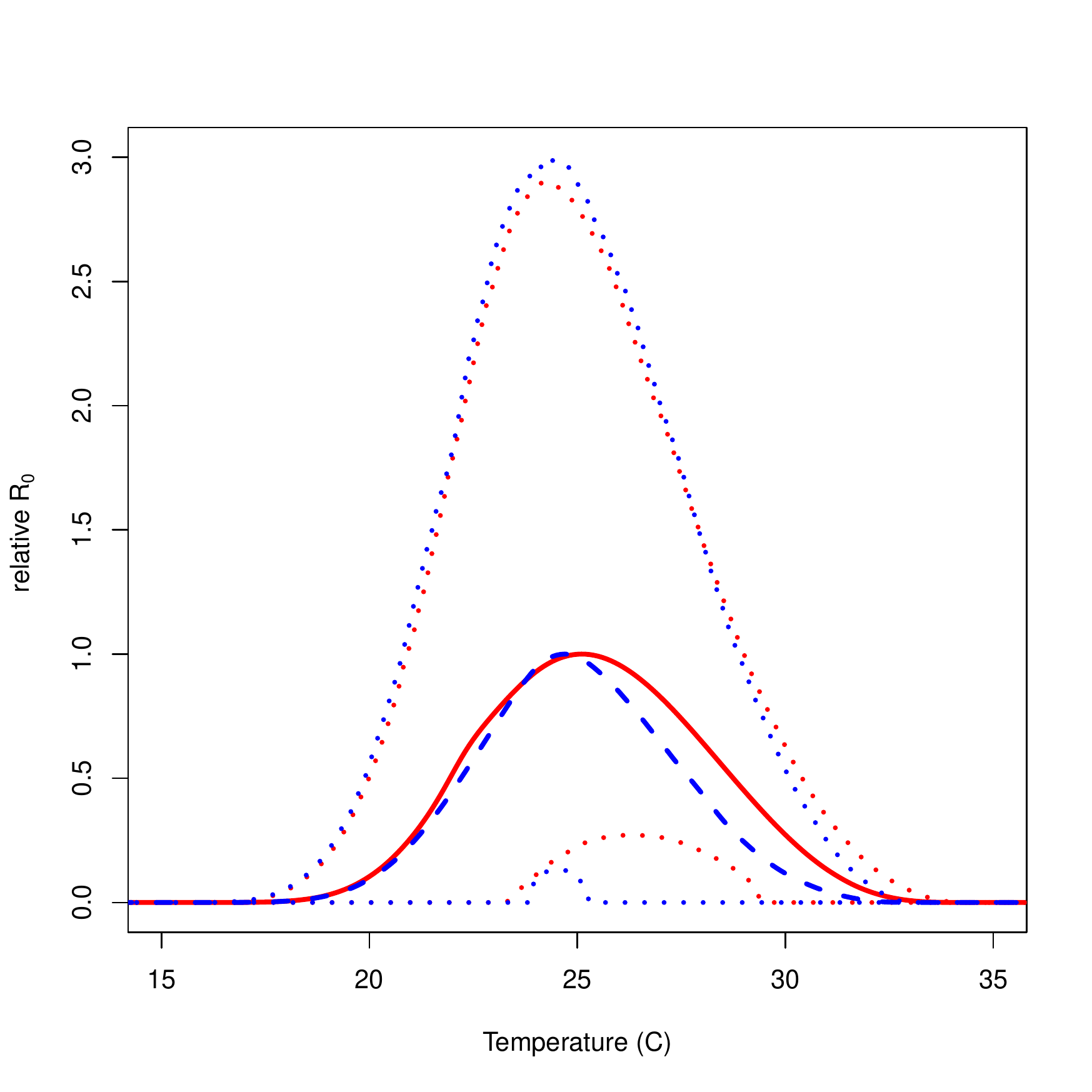}\\
\includegraphics[trim=0 0 0 0, scale=0.5]{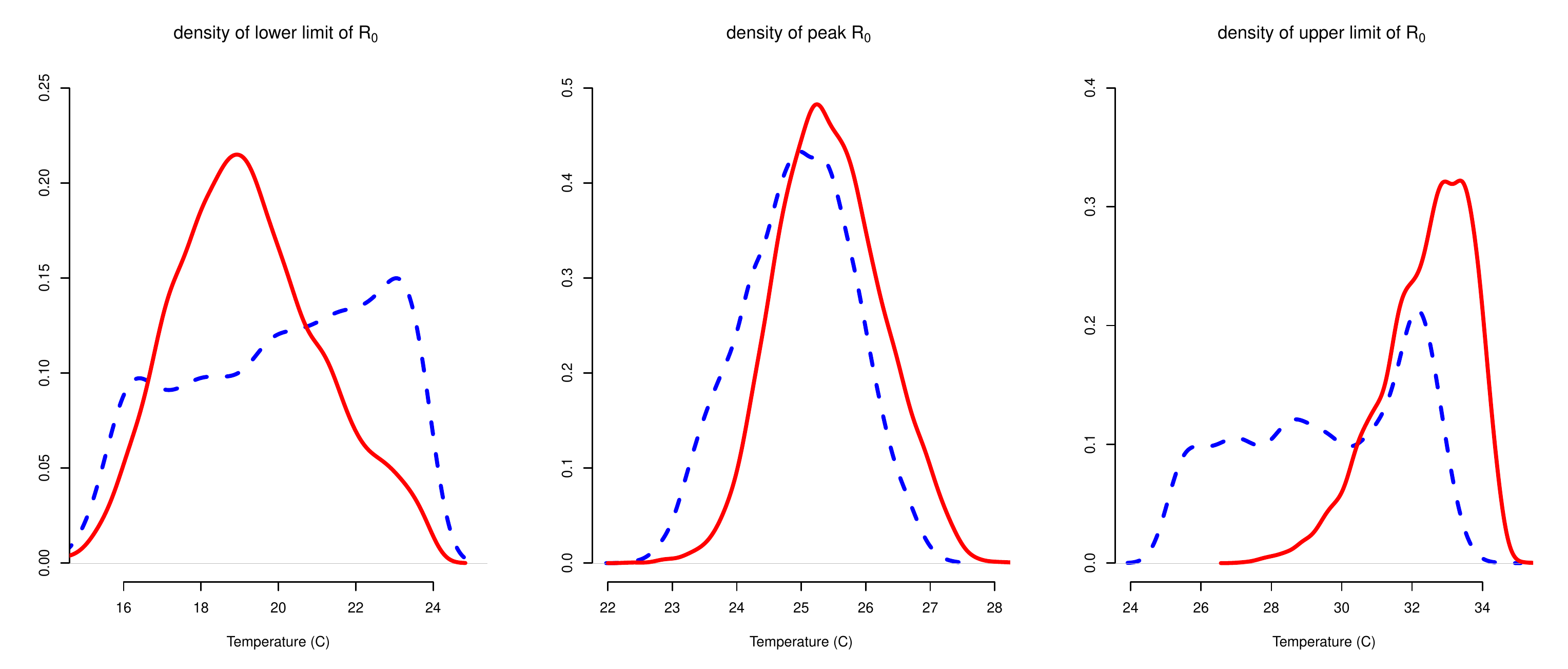}\\
\end{tabular}
\caption{(TOP) Relative $R_0$ ($R_0$ divided by the maximum value of the posterior mean of $R_0$) assuming a quadratic function for vector competence, with uninformative priors for all components (blue, dashed) and informative priors for all components (red, solid). 95\% HPD around each curve are shown as dotted lines. (BOTTOM) Smoothed posterior distributions of the (left) lower temperature limit of $R_0$,  (middle) peak temperature of $R_0$,  (right) upper temperature limit of $R_0$ all assuming a quadratic function for vector competence. Case with uninformative prior is shown as a blue dashed line and with informative prior as a solid red line. The height of the distribution indicates the relative probability of the value of the quantity of interest. }
\label{f:R0summary}
\end{figure}

In Figure \ref{f:R0summary} (top) we show the posterior mean of the temperature dependence of $R_0$ and 95\% HPD
intervals of the temperature response of $R_0$ when both informative and uninformative priors are used. All curves are scaled to the maximum value of the mean $R_0(T)$ curve. These are generated using posterior samples from all components, and so indicate the overall uncertainty in the response curve due to uncertainty in all components, simultaneously. Notice that the mean
$R_0$ curves obtained using default and informative priors are very similar. Further, the
upper 95\% HPD intervals were nearly identical. However, the lower HPD intervals of
$R_0$, especially at higher temperatures, differ considerably as the
additional prior information allowed us to pin down the high temperature transmission limit more
precisely. This can be seen more clearly by looking at the posterior
distributions of the upper and lower temperature limits of  $R_0$ and the distribution of the temperature at peak $R_0$ (Figure \ref{f:R0summary}, bottom). With the default priors, almost the full range of possibilities for the lower (from 0 to 24 $^\circ$C) and upper (from 25 to 45 \degC) limits were equally represented. Adding in prior information indicates support for a slightly lower temperature limit to malaria transmission while the upper limit is at a slightly higher temperature than was predicted with default priors. In other words, the climate envelope where transmission may be possible is slightly larger than would be inferred without prior data. Further, our estimates of the temperatures that can exclude malaria, particularly at the upper end, are more precise. However, the prediction of the temperature of peak transmission, which corresponds to temperatures at which malaria is expected to be most severe and difficult to control, was robust.

As mentioned in the previous section, the most appropriate functional response to describe the temperature dependence of vector competence ($bc$, a compound trait) was ambiguous. Since both functional forms fit the available data well, we examined how using each impacted the posterior inferences for $R_0$. To do this, we calculated $R_0$ using first the posterior samples for the Briere response for $bc$ and then the quadratic. All other components of $R_0$ were allowed to take all possible values of their posterior distributions. Thus our comparison shows how the uncertainty in the functional form for $bc$ impacts the overall uncertainty in $R_0$ given the full uncertainty in the other parameters. The choice of the functional form for $bc$ had little impact on the  the posterior distribution of $R_0(T)$ except at the high temperature limit (Supp.~App.~B). As with other components examined, the vector competence fit with a Briere function exhibited reduced uncertainty in the upper limit compared to the quadratic fit, and as a result decreased the uncertainty in $R_0$ at this limit. Since there is no {\it a priori} reason to prefer one or the other of these, in all further analyses we assumed the quadratic fit, as this resulted in the most uncertainty in $R_0$ at the upper limit, and was thus a more conservative choice. 

\vspace{-0.5cm}
\subsection{{\small Uncertainty in $R_0$ and its sensitivity to temperature, by component}}
\noindent {\bf Which mosquito and parasite traits drive the uncertainty in $R_0$, across temperatures?} 
To answer this question, we used a variation of a traditional sensitivity analysis. We set all but a focal component to its posterior mean.  Then the posterior samples of the focal component were used to calculate the width of the 95\% HPD at each temperature due to only the variation in this single component. We then normalized this to the width of the 95\% HPD for the full posterior of $R_0$ (i.e., when all components were allowed to vary) to approximate the proportion of the uncertainty each component contributed to the full uncertainty in $R_0$.

\begin{figure}[h!]
\centering
\begin{tabular}{cc}
\includegraphics[trim=0 0 0 0, scale=0.45]{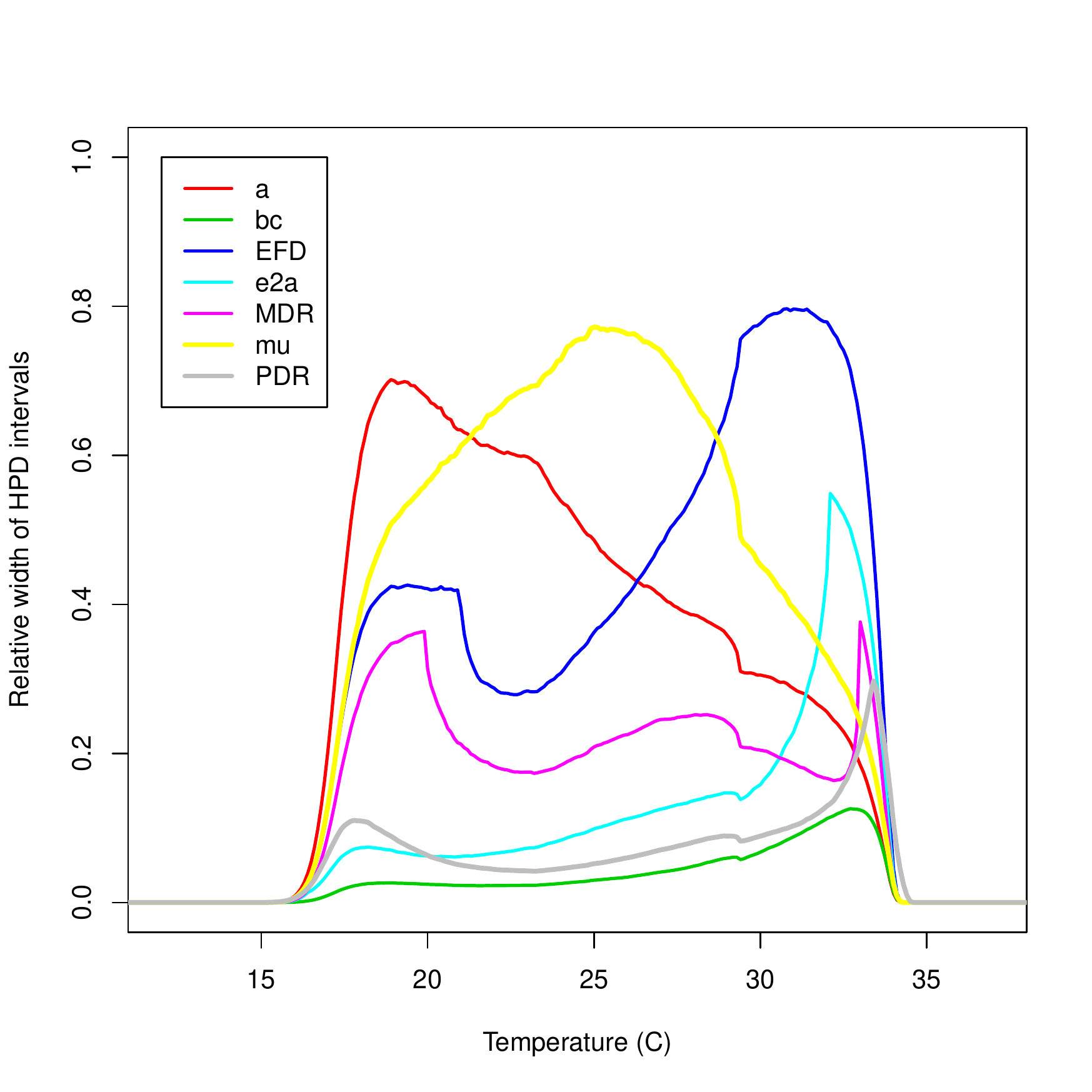}&
\includegraphics[trim=0 0 0 0, scale=0.45]{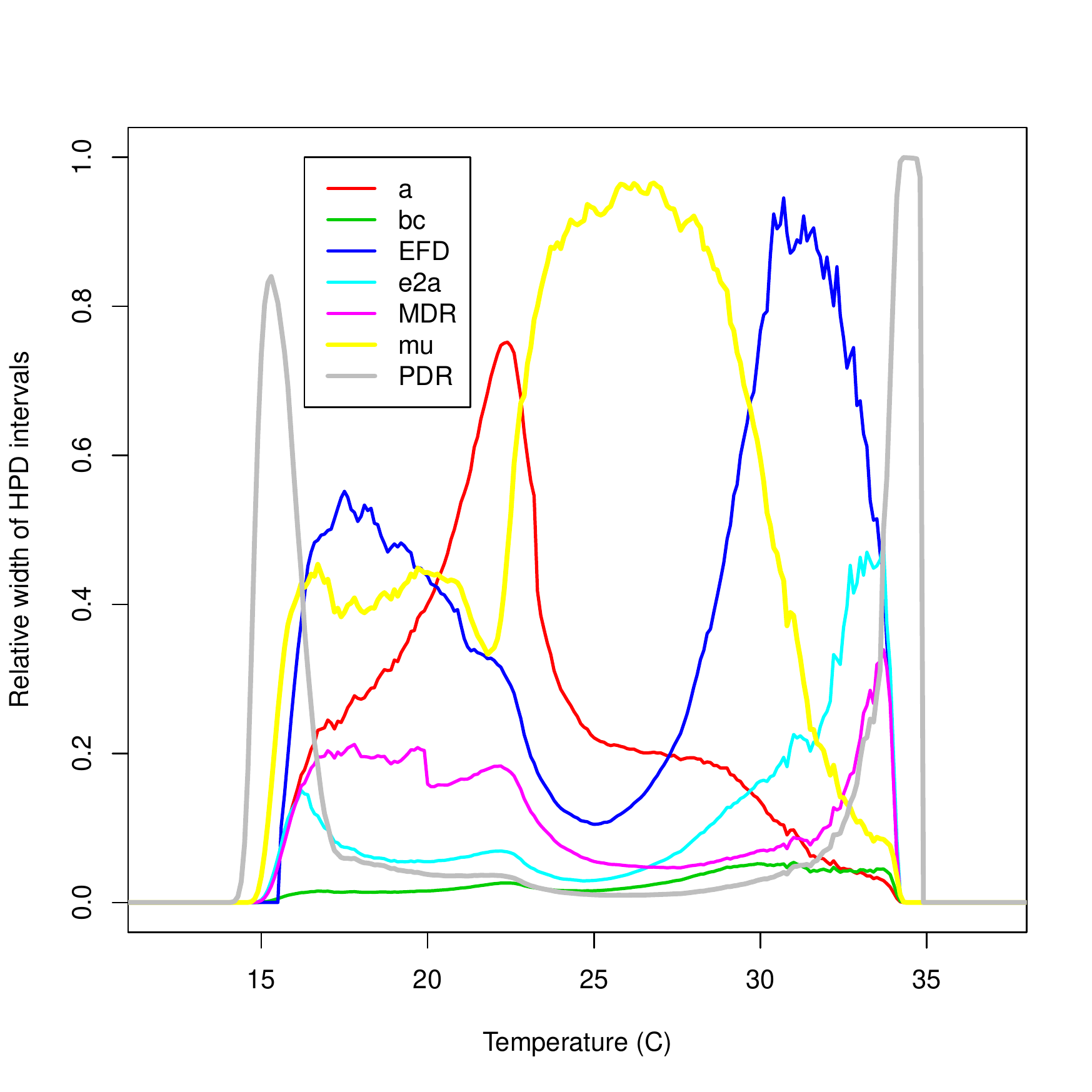}\\
(a) & (b) \\
\end{tabular}
\caption{
(a) Relative width of the 95\% HPD intervals due to uncertainty in each component, compared to uncertainty in $R_0$ overall. Each curve was obtained as follows. For each component, $R_0$ was calculated for the thinned posterior samples of that component, with all other components set to its posterior mean. Then the width of the inner 95\% HPD was calculated at each temperature. This was then normalized to the width of the HPD of the full posterior distribution of $R_0$ (incorporating the full posterior samples from all components simultaneously) at each temperature. (b) Relative width of the 95\% HPD in $\left( \dd{R_0}{T}\right)_\theta$ scaled by the width of the 95\% HPD for $\frac{dR_0}{dT}$ at each temperature, calculated as in (a). In both, a quadratic response for vector competence ($bc$) was used.}
\label{f:R0sensitivity}
\end{figure}

In Figure \ref{f:R0sensitivity} (a) we show the amount of uncertainty in $R_0$ due to a single component, compared to the uncertainty overall. The adult mosquito mortality rate $\mu$, dominated at intermediate temperatures, i.e., the region in which temperature promotes transmission, because $R_0 \propto 1/\mu^3$ (by definition from Equations (\ref{eq:R0}) and (\ref{eq:M})), and at intermediate temperatures the rate of adult mosquito mortality is low. More surprising was the relatively narrow range of temperatures where it dominated: $\mu$ dominated in the middle third of the transmission range, and is primarily determined the height and location of peak $R_0$. This explains why the informative priors had so little impact on the upper HPD interval for $R_0$ (Figure \ref{f:R0summary}) -- the posterior of $\mu$ was not impacted by additional prior information and this component determined the height of the curve. By contrast, the uncertainty on the lower and upper limits for transmission was driven by bite rate and fecundity, respectively. In particular, uncertainty in the bite rate ($a$) drove uncertainty in the lower temperature boundary, while the fecundity ($EFD$), was most important at the upper temperature limit. These two components were also the most important sources of uncertainty, after $\mu$, at intermediate temperatures. Other components, such as vector competence ($bc$), contributed very little to the overall uncertainty, despite the fact that the data on these components were sparse. This was primarily because parameters like this are probabilities, bounded between zero and one, and only impact $R_0$ proportionally. Other  parameters have the potential to vary over orders of magnitude, and thus swamp the impact of these parameters over most of the range of $R_0$. Only when the magnitude of the other parameters are small, such as the temperature extremes, can these contribute significantly to the overall uncertainty.

\noindent {\bf Which mosquito and parasite traits determine the sensitivity of $R_0$ to temperature and the uncertainty in the sensitivity, across temperatures?}  
Next we examined how sensitive $R_0$ is to changes in temperature, that is, how various components contribute to the {\em change} in $R_0$ with temperature, as measured by the derivative of $R_0$ with respect to temperature. We especially focused on the uncertainty of the sensitivity, measured by the HPD interval around the mean sensitivity, and which components drive the uncertainty. We focus on these as they indicate the data that can best help improve our understanding of what determines the shape of $R_0$ across temperatures. 
We started with a standard sensitivity analysis, calculating the derivative of $R_0$ with respect to temperature overall and for each component separately, $\left( \dd{R_0}{T}\right)_\theta$ (see Supp.~App.~A Section  5 for equations). As with the previous analysis, we then set all components, save a focal component to its mean, and used the posterior samples from that focal component to obtain the marginal posterior distribution for the sensitivity. We repeated this for all components, then examined the median sensitivity and the width of the 95\% HPD intervals for the component-wise sensitivity, $\left( \dd{R_0}{T}\right)_\theta$, relative to the overall sensitivity, $\dd{R_0}{T}$ (calculated using the posterior samples from all the components at once) to see which components were driving the sensitivity of $R_0$ to temperature and the uncertainty in the sensitivity. 

In Figure \ref{f:R0sensitivity} (b) we show the uncertainty in the sensitivity of $R_0$ to temperature by each component scaled by the overall uncertainty in the sensitivity of $R_0$ to temperature. For instance, the median sensitivities, by component, indicate that at low temperatures $R_0$ was very sensitive to the bite rate and at high temperatures (even the very highest) the mortality rate drove the response of $R_0$ to temperature (Supp.~App.~C). However, at very high and very low temperatures, other components besides bite rate and mortality began to be important as well, and these other components were less certain. Thus, at the temperature extremes determining where transmission is not possible, the {\em uncertainty} in how sensitive $R_0$ is to temperature was driven by other components, such as fecundity (EFD) and the parasite development rate (PDR).

\vspace{-0.5cm}
\section{Discussion}

Using a Bayesian approach, we identified component traits that were the main sources of uncertainty in $R_0$, and in how sensitive $R_0$ is to temperature. Overall, uncertainty about the temperature limits on transmission was greater than the uncertainty in the optimal temperature for transmission. We found that much of the uncertainty in $R_0$ was due to adult mosquito mortality, $\mu$, as one would expect given $R_0$'s nonlinear dependence on mortality. This contribution was focused in the region of temperatures that promote transmission, where $\mu$ and its uncertainty were small. Other components determined the uncertainty near the temperature limits of $R_0$ in terms of the relative width of the HPD intervals. In particular, near the low temperature limit, the uncertainty in $R_0$ was largely due to uncertainty in the bite rate, $a$, whereas near the high temperature limit the uncertainty was primarily due to uncertainty in fecundity (EFD). Fecundity also contributed a relatively large amount of uncertainty across temperatures. The uncertainty in the high temperature limit itself was determined primarily by the parasite development rate (PDR). \\
\bigskip
The most important empirical data needed to improve model certainty depends on the goals. To resolve the uncertainty in the temperature optimum, empirical studies should focus on measuring adult mosquito mortality rate from 20-30$^\circ$ C, as well as bite rate and fecundity. By contrast, resolving uncertainty surrounding the lower and upper temperature limits on transmission requires measuring 1) bite rate from 15-25$^\circ$ C; 2) fecundity across temperatures, but especially $\sim$25-32$^\circ$ C; and 3) PDR at $\sim$15-16$^\circ$ C and 33-35$^\circ$ C. This last is especially interesting, because PDR is not driving the overall uncertainty at any temperature. Instead, our uncertainty in how {\em sensitive} $R_0$ is to temperature depends on PDR at both temperature extremes. More specifically, our uncertainty in the temperature at which $R_0$ changes from zero is driven by PDR,. This also suggests that PDR component could be determining the temperature limits for malaria transmission, and the  ability of the parasite to evolve at the edges of its thermal limits could determine where malaria could occur. Resolving the temperature limits is particularly important given that warming is expected to expand transmission into currently or recently unsuitable highland areas \citep{siraj_altitudinal_2014}, and may force currently warm, suitable lowland areas above the upper transmission limit. Differences in the presumed thermal limits that inhibit transmission can impact predictions of when and where transmission is likely to occur both now and in the future.  \citep{ryan_climate_2014}. A better understanding of the uncertainty in the temperatures that inhibit transmission should help inform policy priorities as climate changes.

Further, our results provide guidance as to which components may not be as high priority for further work. For instance, our analysis indicates which components are contributing relatively little to the uncertainty in $R_0$, such as vector competence ($bc$). Although these components are necessary for transmission (incompetent vectors cannot transmit disease, for example), investment in reducing uncertainty in these components will have a comparatively small impact on our overall understanding of $R_0$. Thus, these components could be given a lower priority for further empirical effort, especially if these components are difficult or expensive to measure.\\

\bigskip

Our method is unusual in that it combines concepts and approaches from traditional global and local sensitivity analyses with a Bayesian method of inference. Bayesian analyses often focus on the inference and uncertainty aspects, and, for simpler models, model and variable selection. If sensitivity is addressed, it is typically the sensitivity of the posterior to the prior specification. Conversely, most sensitivity analyses would typically be conducted using parameter values selected at random or evenly over some ``reasonable range'' of parameters for a system instead of performing parameter inference as part of the procedure. This traditional approach would be most similar to performing our analysis using samples from informative priors. The choice to take the extra step and fit data before doing the uncertainty and sensitivity analyses depends on whether or not appropriate, high quality data are available for the analysis. If this is not the case, then incorporating additional data could lead to bias in results. However, even moderately good data, such as we have in our example, allowed us to narrow down the regions of parameter space that are reasonable for our system by allowing the data to inform us of how the parameters that describe our thermal response curves are correlated for each component of $R_0$, which is not easy to obtain {\it a priori}.  

On the other hand, in this example, we have been forced to use data from alternative species, even for the focal data. For instance, we used fecundity data for {\it Aedes albopictus} for the focal data, even though we expect these are likely to be different from any {\it Anopheles} species. We also used {\it P.~vivax} data for vector competence. Thus, we may have underestimated the overall uncertainty in $R_0$ and the uncertainty due to these components. It is unclear how to explicitly incorporate this into the current (or any other) framework to quantify the impact of using these data. However, the types of sensitivity analyses we conducted take into account the structure of the model, and give information about how much each component could potentially contribute to the overall uncertainty. We can, therefore, use these analyses to complement our intuition about how to focus research efforts. Thus, although fecundity for Anophelene species and vector competence for {\it P.~falciparum} could be useful (since we had to use alternative species for these analyses), when combined with the rest of the analyses, we would prioritize fecundity as uncertainty in that component has a bigger impact on overall uncertainty.   Further, we might expect that some components that contribute to $R_0$ have biological reasons to be more uncertain, for instance because there is more individual variability in some traits or local adaptation \citep{sternberg_local_2014}. With appropriate data we could explicitly include individual (or population) 
level variation as part of the analysis, or combine data across different populations while allowing differences in thermal response curves within the framework.

Although we applied our methods to a particular formulation of $R_0$ for malaria, the approach is appropriate for mechanistic models of other systems, including other vector-borne disease and species distribution models. To best understand the full uncertainty for a particular disease or species distribution, consideration of multiple mechanistic models and environmental drivers would allow researchers to further understand model uncertainty, and the robustness of predictions to other formulations. For instance, alternative dynamical models could change the estimate of the peak and range of species distributions as the functional relationship between various traits or vital rates and population performance could be substantially different. This is an issue that is largely ignored in the literature addressing disease incidence, species distributions, and climate envelopes. This source of model uncertainty needs to be addressed to fully quantify uncertainty in the response of populations to climate change. We suspect that for many species and diseases, including malaria, the uncertainty inherent in the individual components would often swamp model uncertainty in big picture quantities like $R_0$. This is especially true when data from a focal species are not available at all, as discussed above, and further emphasises the importance of high quality data for parameterising these kinds of models.  However, alternative formulations incorporate each component in different ways, so the conclusions about which components drive uncertainty are likely to be less robust. Consideration of multiple models can indicate what data acquisition should be prioritized, for instance if it is important across a variety of model formulations, and further, which new data would allow better discrimination between competing models. The Bayesian approach allows direct comparison of models and their uncertainty. Thus it has the potential to be a useful tool for identifying concrete recommendations for future research to improve predictions of how factors such as climate change could impact the distribution of malaria and other vector-borne diseases. 

\noindent {\large {\bf Acknowledgements}:}
This work was conducted as part of the Malaria and Climate Change Working Group supported by the Luce Environmental Science to Solutions Fellowship and the National Center for Ecological Analysis and Synthesis, a Center funded by NSF (EF-0553768), the University of California, Santa Barbara and the State of California.  EAM was funded by an NSF Doctoral Dissertation Improvement Grant (DEB-1210378), and an NSF Postdoctoral Research Fellowship (DEB-1202892). Any use of trade, product or firm names in this publication is for descriptive purposes only and does not imply endorsement by the US government.

\bibliographystyle{abbrvnat}
\bibliography{../../../tex/biology,../../../tex/generalbib,../bookchapter/malaria}

\vspace{0.5in}

{\bf Supplementary Appendix A}: Data sources, functional forms, default and informative prior details, equations for sensitivity analysis.\\
\bigskip
{\bf Supplementary Appendix B}: Plots for full results for all components of $R_0$.\\
\bigskip
{\bf Supplementary Appendix C}: Further results for $R_0$ and sensitivity to temperature.\\
\bigskip
{\bf Supplementary Files}: Compiled data, source code for analyses, and output files to reconstruct the analysis described in the manuscript.

\clearpage
\renewcommand{\arraystretch}{2}
{
\begin{table}[ht!] 
\begin{center}
\begin{tabular}{| c |  l |  c |}
  \hline
  Parameter & Definition & Functional Form  \\
  \hline
$a$ & bite rate (1/ gonotrophic cycle length) & Briere \\
MDR & mosquito development rate & Briere \\
$p_{EA}$ & egg to adult survival  & quadratic \\
EFD & fecundity (eggs/female/day) & quadratic \\
$\mu$ & mosquito mortality rate & quadratic \\
$bc$ & vector competence & quadratic (and Briere)\\
PDR & parasite development rate (1/EIP) &  Briere \\
 \hline
\end{tabular} 
\caption[]{Component parameters of $R_0$ and their definitions. 
\label{tb:params}}
\end{center}
\end{table}
}

\clearpage

Figure 1: Posterior mean (solid line) and 95\% credible interval (dashed lines) of the thermal responses for all components of $R_0$, with informative priors together with the main data. Traits modeled with a Briere thermal response  ($cT(T-T_0)\sqrt{(T_m-T)}$) are grouped in the top row, concave-down quadratic ($f(T)=a(T-T_0)(T-T_m)$) in the middle row, and concave-up quadratic ($aT^2+bT+c$) in the bottom row. Data symbols correspond to the species of mosquito or parasite used for the analysis. $\bullet$: {\it An.~gambiae} or {\it P.~falciparum} in {\it An.~gambiae}; $+$: other Anophelene species or {\it P.~falciparum} in other Anophelene species; $\times$: {\it Aedes} species; $\circ$: {\it P.~vivax} in other Anophelene species. \\
\bigskip
Figure 2: (TOP) Relative $R_0$ ($R_0$ divided by the maximum value of the posterior mean) assuming a quadratic function for vector competence, with uninformative priors on all components (blue, dashed) and informative priors on components (red, solid). 95\% HPD around each curve are shown as dotted lines. (BOTTOM) Smoothed posterior distributions of the (left) lower temperature limit of $R_0$,  (middle) peak temperature of $R_0$,  (right) upper temperature limit of $R_0$ all assuming a quadratic function for vector competence. Case with uninformative prior is shown as a blue dashed line and with informative prior as a solid red line. \\ 
\bigskip
Figure 3: (a) Relative width of the 95\% HPD intervals due to uncertainty in each component, compared to uncertainty in $R_0$ overall. Each curve was obtained as followed. For each component, $R_0$ was calculated for the thinned posterior samples of that component, with all other components set to its posterior mean. Then the width of the inner 95\% HPD was calculated at each temperature. This was then normalized to the width of the HPD of the full posterior distribution of $R_0$ at each temperature. (b) Relative width of the 95\% HPD in $\left( \dd{R_0}{T}\right)_\theta$ scaled by the width of the 95\% HPD for $\frac{dR_0}{dT}$ at each temperature, calculated as in (a). In both, a quadratic response for vector competence (bc) was used.

\end{document}